\documentclass[12pt]{article}
\usepackage{graphicx}
\usepackage{epstopdf}
\advance\textheight by \topskip
\begin{document}


\newcommand{\HPA}[1]{{\it Helv.\ Phys.\ Acta.\ }{\bf #1}}
\newcommand{\AP}[1]{{\it Ann.\ Phys.\ }{\bf #1}}
\newcommand{\be}{\begin{equation}}
\newcommand{\ee}{\end{equation}}
\newcommand{\br}{\begin{eqnarray}}
\newcommand{\er}{\end{eqnarray}}
\newcommand{\ba}{\begin{array}}
\newcommand{\ea}{\end{array}}
\newcommand{\bi}{\begin{itemize}}
\newcommand{\ei}{\end{itemize}}
\newcommand{\bn}{\begin{enumerate}}
\newcommand{\en}{\end{enumerate}}
\newcommand{\bc}{\begin{center}}
\newcommand{\ec}{\end{center}}
\newcommand{\ul}{\underline}
\newcommand{\ol}{\overline}
\def\l{\left\langle}
\def\r{\right\rangle}
\def\as{\alpha_{s}}
\def\ycut{y_{\mbox{\tiny cut}}}
\def\yij{y_{ij}}
\def\epem{\ifmmode{e^+ e^-} \else{$e^+ e^-$} \fi}
\newcommand{\eeww}{$e^+e^-\rightarrow W^+ W^-$}
\newcommand{\qqQQ}{$q_1\bar q_2 Q_3\bar Q_4$}
\newcommand{\eeqqQQ}{$e^+e^-\rightarrow q_1\bar q_2 Q_3\bar Q_4$}
\newcommand{\eewwqqqq}{$e^+e^-\rightarrow W^+ W^-\ar q\bar q Q\bar Q$}
\newcommand{\eeqqgg}{$e^+e^-\rightarrow q\bar q gg$}
\newcommand{\eeqloop}{$e^+e^-\rightarrow q\bar q gg$ via loop of quarks}
\newcommand{\eeqqqq}{$e^+e^-\rightarrow q\bar q Q\bar Q$}
\newcommand{\eewwjjjj}{$e^+e^-\rightarrow W^+ W^-\rightarrow 4~{\rm{jet}}$}
\newcommand{\eeqqggjjjj}{$e^+e^-\rightarrow q\bar 
q gg\rightarrow 4~{\rm{jet}}$}
\newcommand{\eeqloopjjjj}{$e^+e^-\rightarrow q\bar 
q gg\rightarrow 4~{\rm{jet}}$ via loop of quarks}
\newcommand{\eeqqqqjjjj}{$e^+e^-\rightarrow q\bar q Q\bar Q\rightarrow
4~{\rm{jet}}$}
\newcommand{\eejjjj}{$e^+e^-\rightarrow 4~{\rm{jet}}$}
\newcommand{\jjjj}{$4~{\rm{jet}}$}
\newcommand{\qqbar}{$q\bar q$}
\newcommand{\ww}{$W^+W^-$}
\newcommand{\ar}{\rightarrow}
\newcommand{\sm}{${\cal {SM}}$}
\newcommand{\Dir}{\kern -6.4pt\Big{/}}
\newcommand{\Dirin}{\kern -10.4pt\Big{/}\kern 4.4pt}
\newcommand{\DDir}{\kern -8.0pt\Big{/}}
\newcommand{\DGir}{\kern -6.0pt\Big{/}}
\newcommand{\wwqqqq}{$W^+ W^-\ar q\bar q Q\bar Q$}
\newcommand{\qqgg}{$q\bar q gg$}
\newcommand{\qloop}{$q\bar q gg$ via loop of quarks}
\newcommand{\qqqq}{$q\bar q Q\bar Q$}

\def\st{\sigma_{\mbox{\scriptsize t}}}
\def\Ord{\buildrel{\scriptscriptstyle <}\over{\scriptscriptstyle\sim}}
\def\OOrd{\buildrel{\scriptscriptstyle >}\over{\scriptscriptstyle\sim}}
\def\jhep #1 #2 #3 {{JHEP} {\bf#1} (#2) #3}
\def\plb #1 #2 #3 {{Phys.~Lett.} {\bf B#1} (#2) #3}
\def\npb #1 #2 #3 {{Nucl.~Phys.} {\bf B#1} (#2) #3}
\def\epjc #1 #2 #3 {{Eur.~Phys.~J.} {\bf C#1} (#2) #3}
\def\zpc #1 #2 #3 {{Z.~Phys.} {\bf C#1} (#2) #3}
\def\jpg #1 #2 #3 {{J.~Phys.} {\bf G#1} (#2) #3}
\def\prd #1 #2 #3 {{Phys.~Rev.} {\bf D#1} (#2) #3}
\def\prep #1 #2 #3 {{Phys.~Rep.} {\bf#1} (#2) #3}
\def\prl #1 #2 #3 {{Phys.~Rev.~Lett.} {\bf#1} (#2) #3}
\def\mpl #1 #2 #3 {{Mod.~Phys.~Lett.} {\bf#1} (#2) #3}
\def\rmp #1 #2 #3 {{Rev. Mod. Phys.} {\bf#1} (#2) #3}
\def\cpc #1 #2 #3 {{Comp. Phys. Commun.} {\bf#1} (#2) #3}
\def\sjnp #1 #2 #3 {{Sov. J. Nucl. Phys.} {\bf#1} (#2) #3}
\def\xx #1 #2 #3 {{\bf#1}, (#2) #3}
\def\hepph #1 {{\tt hep-ph/#1}}
\def\preprint{{preprint}}

\def\beq{\begin{equation}}
\def\beeq{\begin{eqnarray}}
\def\eeq{\end{equation}}
\def\eeeq{\end{eqnarray}}
\def\a0{\bar\alpha_0}
\def\thrust{\mbox{T}}
\def\Thrust{\mathrm{\tiny T}}
\def\ae{\alpha_{\mbox{\scriptsize eff}}}
\def\ap{\bar\alpha_p}
\def\as{\alpha_{\mathrm{S}}}
\def\aem{\alpha_{\mathrm{EM}}}
\def\b0{\beta_0}
\def\cN{{\cal N}}
\def\cd{\chi^2/\mbox{d.o.f.}}
\def\Ecm{E_{\mbox{\scriptsize cm}}}
\def\ee{e^+e^-}
\def\enap{\mbox{e}}
\def\eps{\epsilon}
\def\ex{{\mbox{\scriptsize exp}}}
\def\GeV{\mbox{\rm{GeV}}}
\def\half{{\textstyle {1\over2}}}
\def\jet{{\mbox{\scriptsize jet}}}
\def\kij{k^2_{\bot ij}}
\def\kp{k_\perp}
\def\kps{k_\perp^2}
\def\kt{k_\bot}
\def\lms{\Lambda^{(n_{\rm f}=4)}_{\overline{\mathrm{MS}}}}
\def\mI{\mu_{\mathrm{I}}}
\def\mR{\mu_{\mathrm{R}}}
\def\MSbar{\overline{\mathrm{MS}}}
\def\mx{{\mbox{\scriptsize max}}}
\def\NP{{\mathrm{NP}}}
\def\pd{\partial}
\def\pt{{\mbox{\scriptsize pert}}}
\def\pw{{\mbox{\scriptsize pow}}}
\def\so{{\mbox{\scriptsize soft}}}
\def\st{\sigma_{\mbox{\scriptsize tot}}}
\def\ycut{y_{\mathrm{cut}}}
\def\slashchar#1{\setbox0=\hbox{$#1$}           
     \dimen0=\wd0                                 
     \setbox1=\hbox{/} \dimen1=\wd1               
     \ifdim\dimen0>\dimen1                        
        \rlap{\hbox to \dimen0{\hfil/\hfil}}      
        #1                                        
     \else                                        
        \rlap{\hbox to \dimen1{\hfil$#1$\hfil}}   
        /                                         
     \fi}                                         %
\def\etmiss{\slashchar{E}^T}
\def\Meff{M_{\rm eff}}
\def\Ord{\lsim}
\def\OOrd{\gsim}
\def\tq{\tilde q}
\def\tchi{\tilde\chi}
\def\lsp{\tilde\chi_1^0}

\def\gam{\gamma}
\def\ph{\gamma}
\def\be{\begin{equation}}
\def\ee{\end{equation}}
\def\bea{\begin{eqnarray}}
\def\eea{\end{eqnarray}}
\def\lsim{\:\raisebox{-0.5ex}{$\stackrel{\textstyle<}{\sim}$}\:}
\def\gsim{\:\raisebox{-0.5ex}{$\stackrel{\textstyle>}{\sim}$}\:}

\def\ino{\mathaccent"7E} \def\gluino{\ino{g}} \def\mgluino{m_{\gluino}}
\def\sqk{\ino{q}} \def\sup{\ino{u}} \def\sdn{\ino{d}}
\def\chargino{\ino{\omega}} \def\neutralino{\ino{\chi}}
\def\cab{\ensuremath{C_{\alpha\beta}}} \def\proj{\ensuremath{\mathcal P}}
\def\dab{\delta_{\alpha\beta}}
\def\zz{s-M_Z^2+iM_Z\Gamma_Z} \def\zw{s-M_W^2+iM_W\Gamma_W}
\def\prop{\ensuremath{\mathcal G}} \def\ckm{\ensuremath{V_{\rm CKM}^2}}
\def\aem{\alpha_{\rm EM}} \def\stw{s_{2W}} \def\sttw{s_{2W}^2}
\def\nc{N_C} \def\cf{C_F} \def\ca{C_A}
\def\qcd{\textsc{Qcd}} \def\susy{supersymmetric} \def\mssm{\textsc{Mssm}}
\def\slash{/\kern -5pt} \def\stick{\rule[-0.2cm]{0cm}{0.6cm}}
\def\h{\hspace*{-0.3cm}}

\def\ims #1 {\ensuremath{M^2_{[#1]}}}
\def\tw{\tilde \chi^\pm}
\def\tz{\tilde \chi^0}
\def\tf{\tilde f}
\def\tl{\tilde l}
\def\ppb{p\bar{p}}
\def\gl{\tilde{g}}
\def\sq{\tilde{q}}
\def\sqb{{\tilde{q}}^*}
\def\qb{\bar{q}}
\def\sqL{\tilde{q}_{_L}}
\def\sqR{\tilde{q}_{_R}}
\def\ms{m_{\tilde q}}
\def\mg{m_{\tilde g}}
\def\Gs{\Gamma_{\tilde q}}
\def\Gg{\Gamma_{\tilde g}}
\def\md{m_{-}}
\def\eps{\varepsilon}
\def\Ce{C_\eps}
\def\dnq{\frac{d^nq}{(2\pi)^n}}
\def\DR{$\overline{DR}$\,\,}
\def\MS{$\overline{MS}$\,\,}
\def\DRm{\overline{DR}}
\def\MSm{\overline{MS}}
\def\ghat{\hat{g}_s}
\def\shat{\hat{s}}
\def\sihat{\hat{\sigma}}
\def\Li{\text{Li}_2}
\def\bs{\beta_{\sq}}
\def\xs{x_{\sq}}
\def\xsa{x_{1\sq}}
\def\xsb{x_{2\sq}}
\def\bg{\beta_{\gl}}
\def\xg{x_{\gl}}
\def\xga{x_{1\gl}}
\def\xgb{x_{2\gl}}
\def\lsp{\tilde{\chi}_1^0}

\def\gluino{\mathaccent"7E g}
\def\mgluino{m_{\gluino}}
\def\squark{\mathaccent"7E q}
\def\msquark{m_{\mathaccent"7E q}}
\def\M{ \overline{|\mathcal{M}|^2} }
\def\utm{ut-M_a^2M_b^2}
\def\MiLR{M_{i_{L,R}}}
\def\MiRL{M_{i_{R,L}}}
\def\MjLR{M_{j_{L,R}}}
\def\MjRL{M_{j_{R,L}}}
\def\tiLR{t_{i_{L,R}}}
\def\tiRL{t_{i_{R,L}}}
\def\tjLR{t_{j_{L,R}}}
\def\tjRL{t_{j_{R,L}}}
\def\tg{t_{\gluino}}
\def\uiLR{u_{i_{L,R}}}
\def\uiRL{u_{i_{R,L}}}
\def\ujLR{u_{j_{L,R}}}
\def\ujRL{u_{j_{R,L}}}
\def\ug{u_{\gluino}}
\def\utot{u \leftrightarrow t}
\def\ar{\to}
\def\sqk{\mathaccent"7E q}
\def\sup{\mathaccent"7E u}
\def\sdn{\mathaccent"7E d}
\def\chargino{\mathaccent"7E \chi}
\def\neutralino{\mathaccent"7E \chi}
\def\slepton{\mathaccent"7E l}
\def\M{ \overline{|\mathcal{M}|^2} }
\def\cab{\ensuremath{C_{\alpha\beta}}}
\def\ckm{\ensuremath{V_{\rm CKM}^2}}
\def\zz{s-M_Z^2+iM_Z\Gamma_Z}
\def\zw{s-M_W^2+iM_W\Gamma_W}
\def\s22w{s_{2W}^2}

\newcommand{\cpmtwo}    {\mbox{$ {\chi}^{\pm}_{2}                    $}}
\newcommand{\cpmone}    {\mbox{$ {\chi}^{\pm}_{1}                    $}}

\begin{flushright}
{SHEP-08-21}\\
\today
\end{flushright}
\vskip0.1cm\noindent
\begin{center}
{\Large {\bf One-loop Electro-Weak Corrections \\[0.25cm]
to Three-jet Observables of $b$-quarks \\[0.45cm] in
$e^+e^-$ Annihilations\footnote{Work supported in 
part by the U.K. Science and Technology Facilities Council 
(STFC),
by the European Union (EU) under contract MRTN-CT-2006-035505 (HEPTOOLS FP6 RTN) and by the 
Italian Ministero dell'Istruzione, dell'Universit\`a e della Ricerca
(MIUR) under contract 2006020509\_004.}}}
\\[1.5cm]
{\large C.M. Carloni-Calame$^{1}$, S. Moretti$^{1}$, F. Piccinini$^{2}$ and D.A. Ross$^{1}$}\\[0.15 cm]
{\it $^1$ School of Physics and Astronomy, University of Southampton}\\
{\it Highfield, Southampton SO17 1BJ, UK}
\\[0.5cm]
{\em $^2$ INFN - Sezione di Pavia,
    Via Bassi 6, 27100 Pavia,
Italy}
\\[0.5cm]
\end{center}

\begin{abstract}
{\small
\noindent
We compute the full one-loop EW
contributions of ${\cal O}(\alpha_{\rm S}\alpha_{\rm{EM}}^3)$ 
entering the electron-positron into two $b$-quarks and one gluon cross section at 
the $Z$ peak and LC energies.
We include both factorisable and non-factorisable virtual corrections,
photon bremsstrahlung but not the real emission of $W^\pm$
and $Z$ bosons. Their importance for the measurement of $\alpha_{\rm S}$ 
from jet rates and shape variables is explained qualitatively and illustrated
quantitatively. Their impact on the forward-backward asymmetry is also analysed.}
\end{abstract}

\section{Introduction}
\label{Sec:Intro}
Jet samples enriched in $b$-quarks produced in $e^+e^-$ annihilations are used
for sophisticated tests of QCD, primarily because they enable one to distinguish between
quarks and gluons, thanks to $b$-flavour tagging (typically, by exploiting high $p_T$ leptons
and/or microvertex techniques). This is unlike the case of lighter flavours\footnote{With the possible
exception of $c$-quarks, whose tagging efficiency is however much lower in comparison to that 
of $b$-quarks, so as to make the former much less suitable  than the latter to phenomenological
investigation.}. To name but a few examples, by studying $b$-jet samples, one can: (i) verify
the flavour independence of $\alpha_{\rm S}$; (ii)
study the properties of the QCD force carrier (the gluon); (iii) measure the $b$-quark mass. 
(For an authoritative review on experimental tests of QCD in $e^+e^-$ events see
\cite{Bethke:2001ih} and references therein.)

Of particular relevance is the three-jet sample, as it impinges on all types of
analysis (i)--(iii). It is therefore of paramount importance to give predictions for the
$e^+e^-\to b\bar b g$ cross section to the highest degree of precision, thereby necessarily
involving the computation of all higher order corrections within the 
Standard Model (SM). 
Whilst the computation of QCD effects through one-loop has been tackled some time ago \cite{BMM,bbgNLO}
the case of Electro-Weak (EW) corrections is not available
in the literature. We remedy this shortcoming here, by calculating  the full
one-loop EW corrections
to $b\bar b g$ observables in electron-positron annihilations
generated via the interference of the graphs
in Figs.~1--6 of Ref.~\cite{poleeEW} (see also \cite{poleeEW1,oldpapers})
with the tree-level ones for $e^+e^-\to \gamma^*,Z\to \bar b bg$. In doing so,
notice that we will be including photon bremsstrahlung;
in contrast, we will refrain from computing real $W^\pm$ and $Z$ boson 
radiation, as we will argue that this may not enter the experimental
jet samples.

Finally, notice that, while QCD corrections are dominant at low $e^+e^-$ energy, EW ones become relatively more and more important
as the latter grows larger, because of surviving Sudakov logarithms from which QCD interactions are immune.
Besides, EW corrections also carry the hallmark of parity-violating effects, which are generally peculiar to
new physics beyond the SM, so that they ought to be accounted for in its quest. 

The plan of the rest of the paper is as follows. In the next Section,
we describe the calculation and give the input parameters and jet-selection algorithms. Then, in Sect.~\ref{Sec:Results},
we present our numerical results for LEP1/SLC, LEP2 and a future Linear Collider (LC). We conclude in Sect.~\ref{Sec:Conclusions}.

\section{Calculation}
\label{Sec:Calculation}

The procedures adopted to carry out our computation have been described in \cite{poleeEW},
to which we refer the reader for the most technical details. Here, we would only like to point out that
we have neglected the masses of the $b$-quarks throughout. However,  whenever there is a $W^\pm$ boson
in the virtual loops, account has to be taken of the mass of the top (anti)quark, which we have done. 
Furthermore,  before proceeding to show our results, we should mention the numerical
parameters used for our simulations. We have
taken the top (anti)quark to have a
mass $m_t = 171.6$ GeV. The $Z$ mass used was $M_Z = 91.18$ GeV and was
related to the $W^\pm$ mass, $M_W$, via the SM formula
$M_W = M_Z \cos \theta_W$, where $\sin^2 \theta_W =$ 0.222478. The $Z$
width was $\Gamma_Z = 2.5$ GeV.
Also notice that, where relevant, Higgs contributions were included
with $M_H=115$ GeV. For the strong
coupling constant, $\alpha_{\rm{S}}$, we have used the two-loop expression with
$\Lambda^{(n_f=4)}_{\rm{QCD}}=0.325$ GeV in the
$\overline{\mathrm{MS}}$ scheme, yielding
$\alpha_{\rm{S}}^{\overline{\mathrm{MS}}}(M_Z^2)=0.118$.

As for the jet definition,  partonic momenta are clustered into jets 
according to the
Cambridge jet algorithm~\cite{cambridge} (e.g., when $y_{ij}<y_{cut}$
with $y_{cut}=0.001$), the jets are required to lie in the central
detector region $30^\circ<\theta_{\mathrm{jets}}<150^\circ$ and we request that
the invariant mass of the jet system $M_{b\bar bg}$ is larger than
$0.75\times\sqrt{s}$.
If a real photon is present in the final state, it is
clustered according to the same algorithm, but we require that
at least three ``hadronic'' jets are left at the end
(i.e., events in which the photon is resolved are rejected)\footnote{As explained in \cite{poleeEW}, this serves a twofold purpose. On the
one hand, from the experimental viewpoint, a resolved (energetic and isolated)
single photon is never treated as a jet. On the other hand, from a theoretical
viewpoint, this enables us to remove divergent contributions appearing 
whenever an unresolved gluon is ejected via an infrared 
(soft and/or collinear) emission, as
we are not computing here ${\cal O}(\alpha_{\rm S}\alpha_{\rm{EM}}^3)$
one-loop QCD contributions to $e^+e^-\to b\bar b\gamma$.}.
We further make the assumption that both $b$-jets can be tagged (as described), including
their charge (e.g., via the emerging lepton or the jet-charge method). (For sake of
illustration we take the efficiency to be one.) In order to show 
the behaviour of the EW corrections we are calculating, other than
scanning in the collider energy, we have considered here the three discrete
values of $\sqrt{s}=M_Z$ (also in view of a GigaZ option of a future LC), 
$\sqrt{s}=350$ (the $t\bar t$ threshold) GeV and $\sqrt{s}=1$ TeV (as representative of
the Sudakov regime).

\section{Numerical Results}
\label{Sec:Results}

\begin{figure}\begin{center}
\includegraphics[width=12cm]{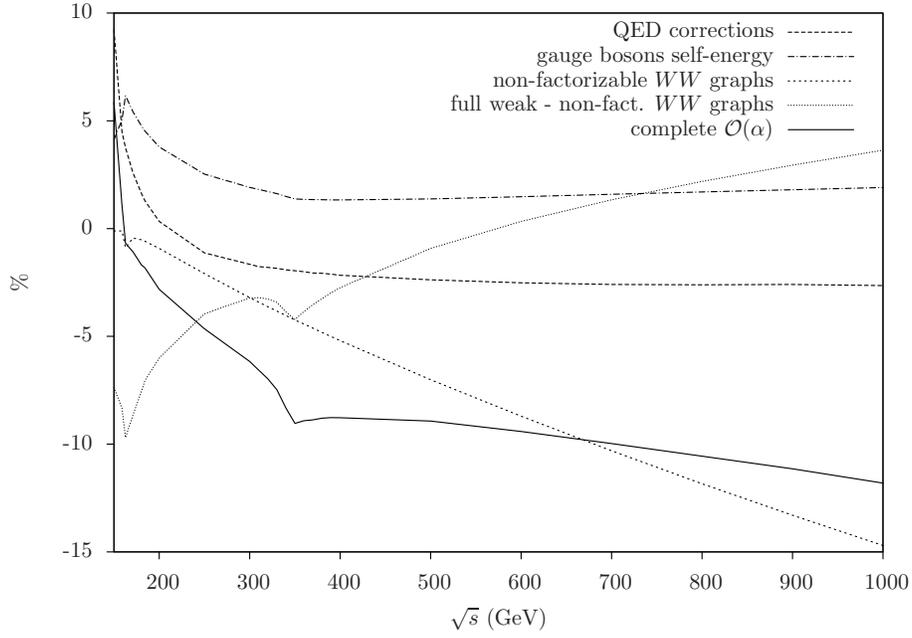}
\caption{Relative effect on the integrated cross section due to
  different contributions to the order $\alpha\equiv
\alpha_{\rm{EM}}$ correction, as a
  function of the CM energy.}
\label{energyscan}\end{center}\end{figure}

In Fig.~\ref{energyscan}, we present
the effects on the cross
section (integrated within the experimental cuts defined in the previous
Section) induced by different terms of the order $\alpha_{\rm S}
\alpha_{\rm{EM}}^3$ contribution relative to the lowest-order cross section for $e^+e^-\to b\bar bg$,
plotted as a function of the Centre-of-Mass
(CM) energy, in
the range from 150 GeV to 1 TeV. The curves represent the effects of
the QED (virtual and real) corrections only, the gauge
bosons self-energy corrections, the non-factorisable graphs involving 
four- and five-point functions with $WW$
exchange\footnote{This is a gauge invariant subset of the complete 
corrections.}, 
the weak corrections with the non-factorizing $WW$ graphs removed
 and the sum of the previous
ones. Notice that 
the total effect is increasingly negative, as $\sqrt s$ gets larger, 
reaching the $-13\%$ or so level at
1 TeV. As already stressed in Ref.~\cite{poleeEW} and visible from the
plot here, such a  big negative correction is mainly due to the presence of the
$WW$ non-factorisable graphs, which develop the aforementioned 
large Sudakov double logarithms in the very high energy regime.
The pattern of the various corrections seen here is not very different from 
the one seen in Ref.~\cite{poleeEW} (Fig. 7 therein) with the notable 
difference that in the case of $b$-samples one can appreciate the onset
of the $t\bar t$ virtual threshold, which was instead invisible in the
fully flavoured sample of Ref.~\cite{poleeEW}.

\begin{figure}\begin{center}
\includegraphics[width=12cm]{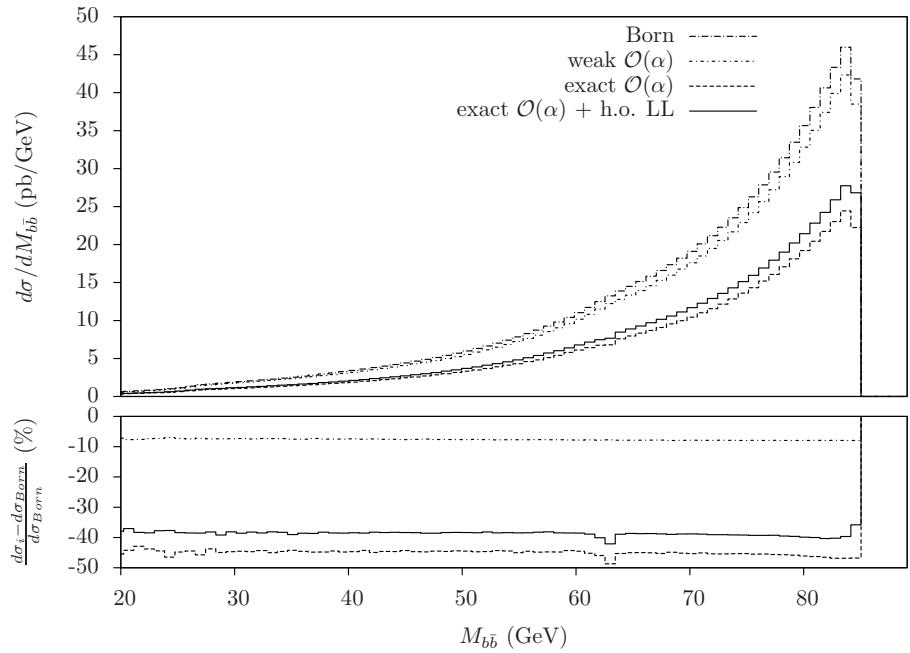}
\caption{$b\bar b$ invariant mass distribution at the $Z$ peak.}
\label{minbbj-peak}\end{center}\end{figure}
\begin{figure}\begin{center}
\includegraphics[width=12cm]{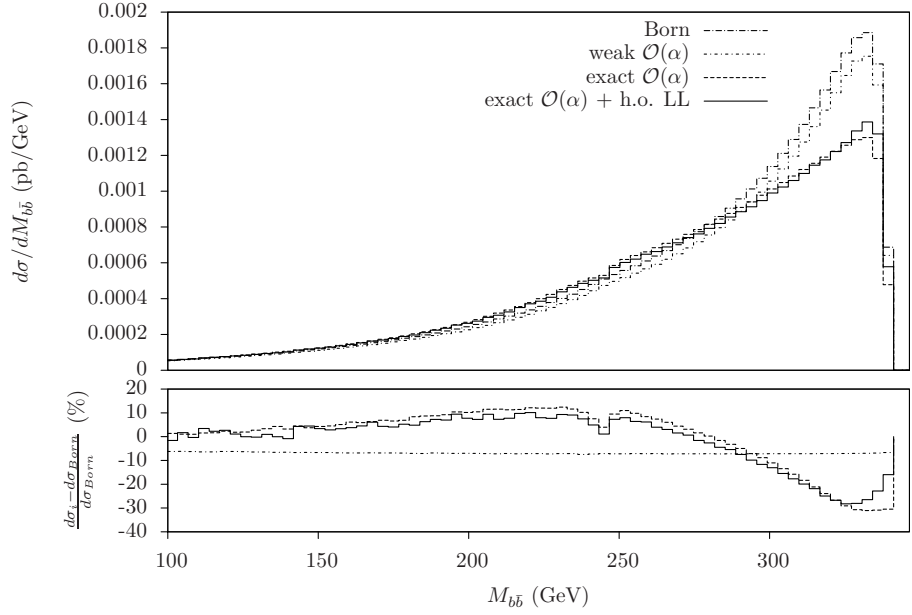}
\caption{$b\bar b$ invariant mass distribution at 350 GeV.}
\label{minbbj-350}\end{center}\end{figure}
\begin{figure}\begin{center}
\includegraphics[width=12cm]{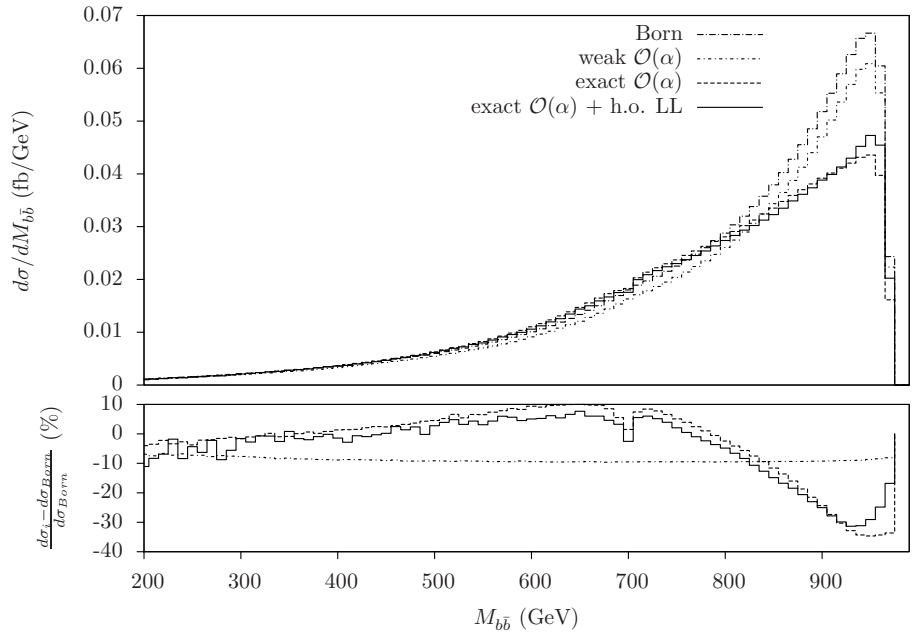}
\caption{$b\bar b$ invariant mass distribution at 1 TeV.}
\label{minbbj-1000}\end{center}\end{figure}
The ability to efficiently tag $b$-quark jets enables one to define
observables in $b\bar bg$ final states
which are not (easily) reconstructable in the case
of the full three-jet sample. One example is the invariant mass
of the $b\bar b$ pair, $M_{b\bar b}$, which we plot in 
Figs.~\ref{minbbj-peak}--\ref{minbbj-1000}.
Here,
the largest contribution to the total correction  comes
from QED Initial 
State Radiation (ISR), primarily because of the radiative return phenomenon.

In fact, the correction is up to
20\% for very high energies and intermediate invariant mass and -30\% for large invariant mass
whereas it is a (constant) $-45\%$ 
at small energy. 
The purely weak
corrections are negative and at the level of $-15\%$
at most, for all energies. The higher order QED
radiation tends to compensate the order $\alpha_{\rm{EM}}$ effect, since it 
enhances the cross section. One should also notice that,
again thanks to a Sudakov effect, by raising the CM energy,
the relative weight of the weak corrections becomes more important
and the effect of higher order QED
corrections diminishes. Finally, it is worth noticing that, far from the
$Z$ peak, the cut $M_{b\bar bg}>0.75\times\sqrt{s}$ is more effective in
suppressing the radiative return phenomenon, reducing in turn the
relative effect of ISR.

\begin{figure}\begin{center}
\includegraphics[width=12cm]{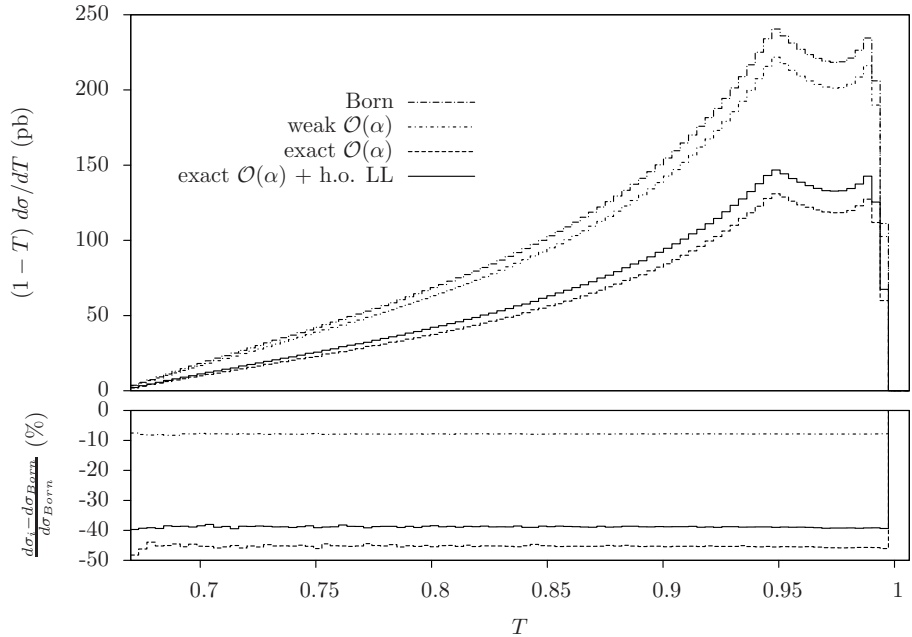}
\caption{$(1-T)\frac{d\sigma}{dT}$ distribution at the $Z$ peak.}
\label{thrust-peak}\end{center}\end{figure}
\begin{figure}\begin{center}
\includegraphics[width=12cm]{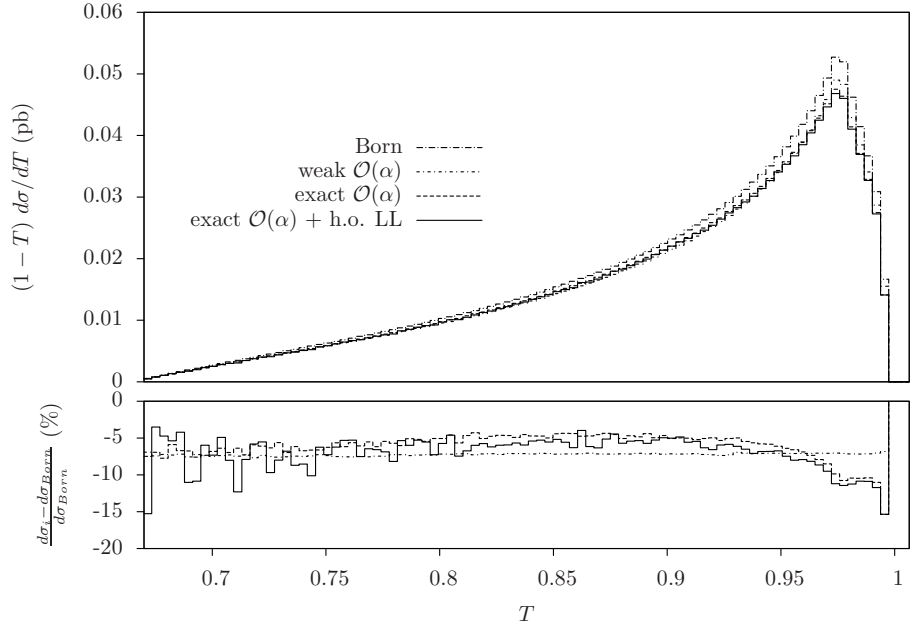}
\caption{$(1-T)\frac{d\sigma}{dT}$ distribution at 350 GeV.}
\label{thrust-350}\end{center}\end{figure}
\begin{figure}\begin{center}
\includegraphics[width=12cm]{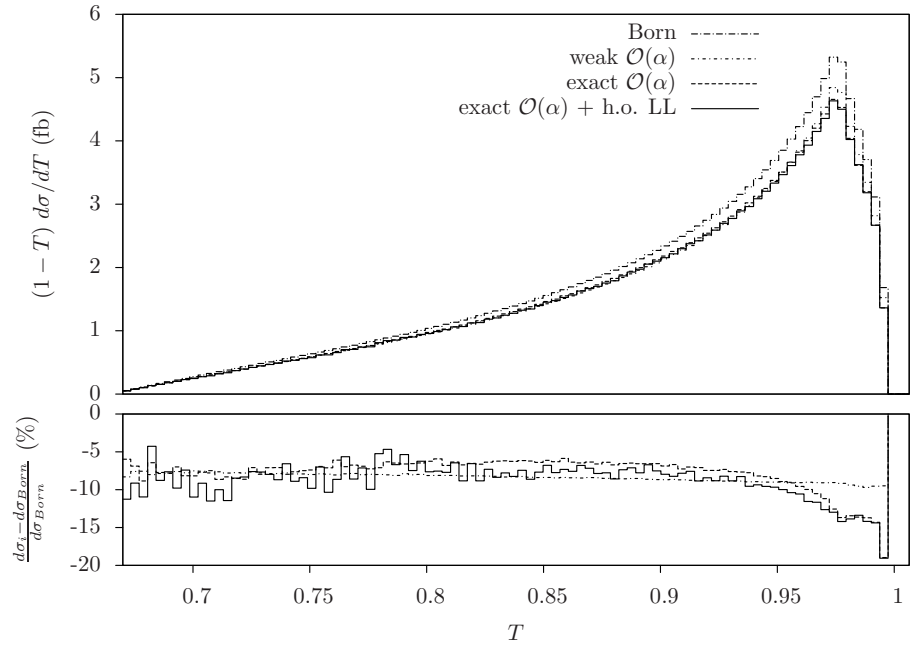}
\caption{$(1-T)\frac{d\sigma}{dT}$ distribution at 1 TeV.}
\label{thrust-1000}\end{center}\end{figure}
In the following some event shape variables are 
considered\footnote{Recalling footnote 3, one should notice that 
these observables will depend on $y_{cut}$. However, this reflects
standard experimental procedures \cite{ALEPH} aiming at removing 
resolved photons from the hadronic sample and
we also have verified that the relative size of the EW effects computed
here does not depend on $y_{cut}$.}: 
the {thrust} $T$~\cite{thrust} and the 
 {\em C-}parameter \cite{cparam} (see Ref.~\cite{kunsztnason} 
for their definitions).  
In Figs.~\ref{thrust-peak},~\ref{thrust-350} and~\ref{thrust-1000},
the spectrum of $(1-T)\frac{d\sigma}{dT}$ is shown. 
This distribution is one
of the key observables used for the measurement of $\alpha_{\rm S}$ 
in $e^+e^-$
collisions~\cite{kunsztnason}. It is worth noticing that whilst the
purely weak corrections give an almost constant effect on the whole
$T$ range, the presence of the real bremsstrahlung gives a non-trivial
effect in the region $T>0.95$ (at least at higher energies). 
The numerical impact 
of the various contributions is similar in the case of the 
{\em C-}parameter. Here, the region where QED real radiation introduces 
non-trivial effects is $C \lsim$~0.1. Given the size of the various EW
corrections (the QED ones being of tens of percent), including the purely
weak ones (which are steadily at --7\% or so) and 
in view of a precise measurement of
$\alpha_{\rm S}$ from $b\bar bg$ samples
at future LCs, it is clear that such effects
will play an important role and thus cannot be neglected.
\begin{figure}\begin{center}
\includegraphics[width=12cm]{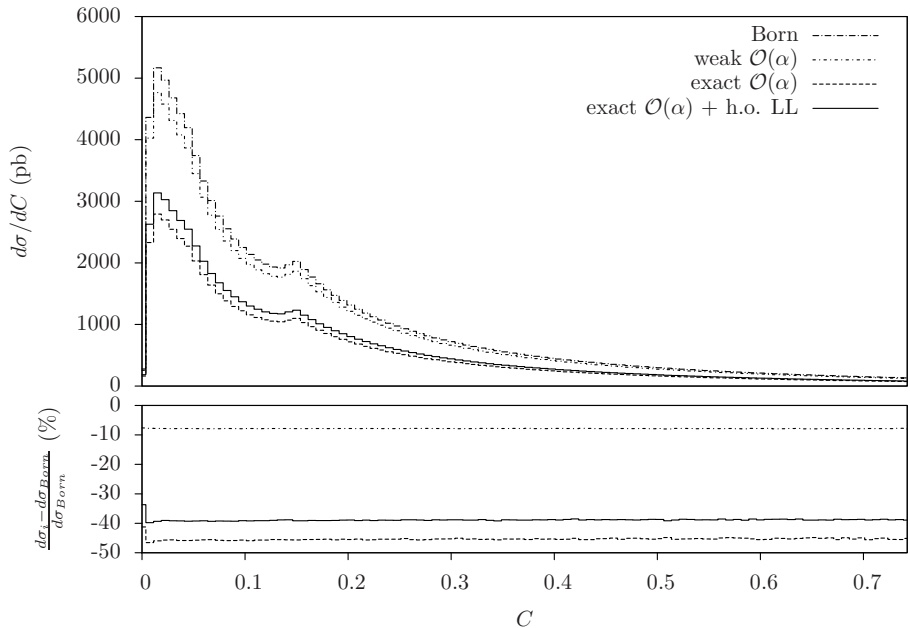}
\caption{$\frac{d\sigma}{dC}$ distribution at the $Z$ peak.}
\label{cpar-peak}\end{center}\end{figure}
\begin{figure}\begin{center}
\includegraphics[width=12cm]{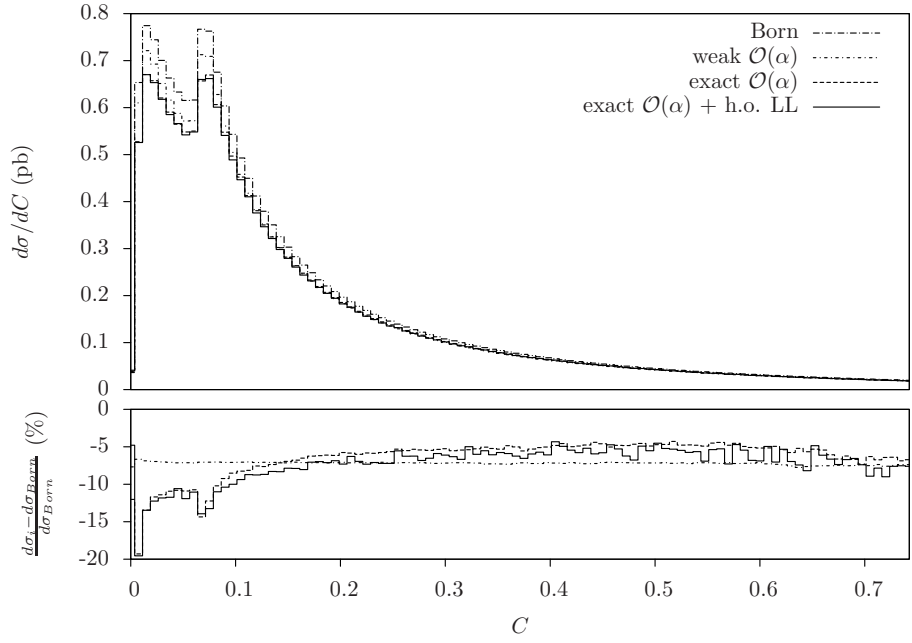}
\caption{$\frac{d\sigma}{dC}$ distribution at 350 GeV.}
\label{cpar-350}\end{center}\end{figure}
\begin{figure}\begin{center}
\includegraphics[width=12cm]{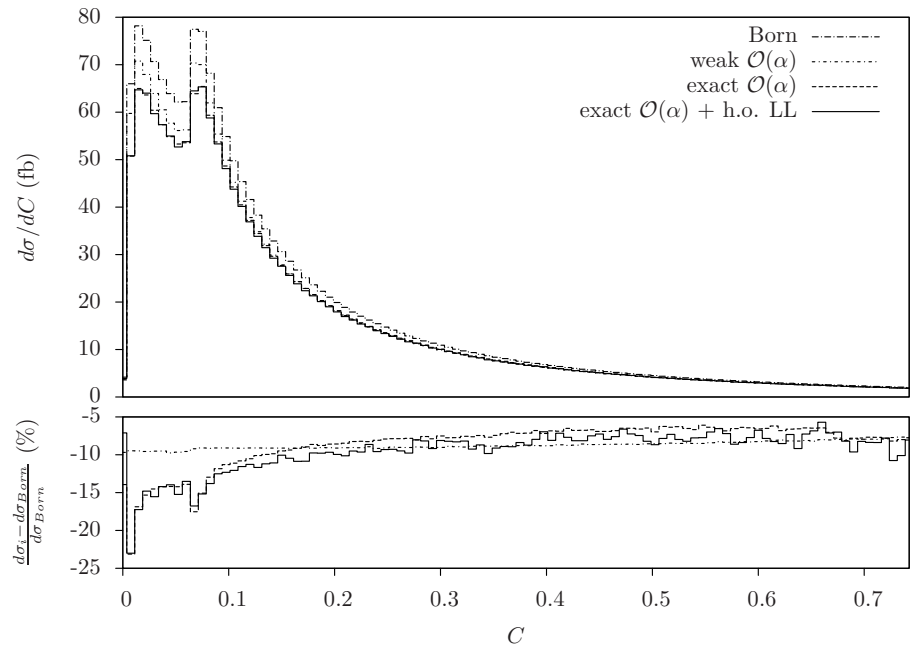}
\caption{$\frac{d\sigma}{dC}$ distribution at 1 TeV.}
\label{cpar-1000}\end{center}\end{figure}

\begin{figure}\begin{center}
\includegraphics[width=12cm]{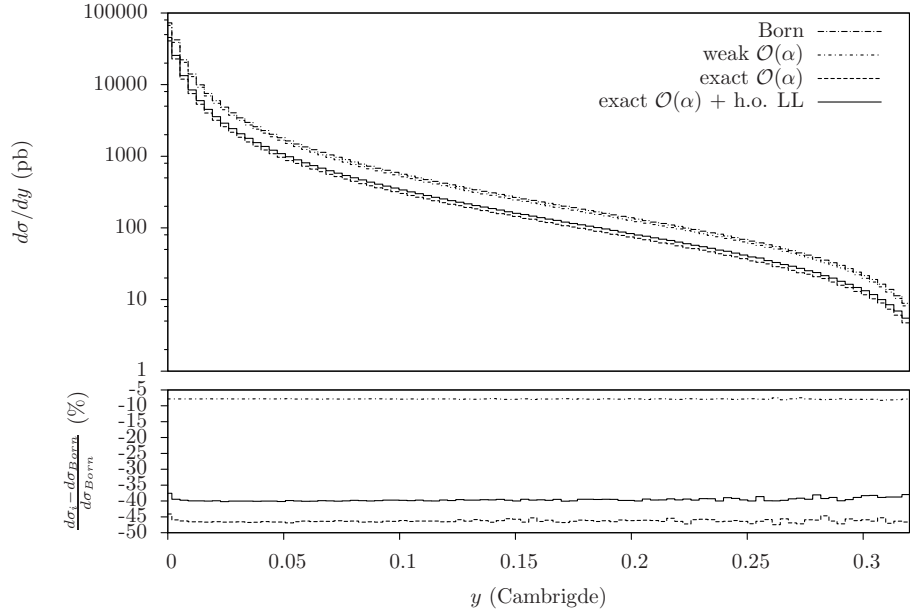}
\caption{Cambridge $y$ distribution at the $Z$ peak.}
\label{ycam-peak}\end{center}\end{figure}

\begin{figure}\begin{center}
\includegraphics[width=12cm]{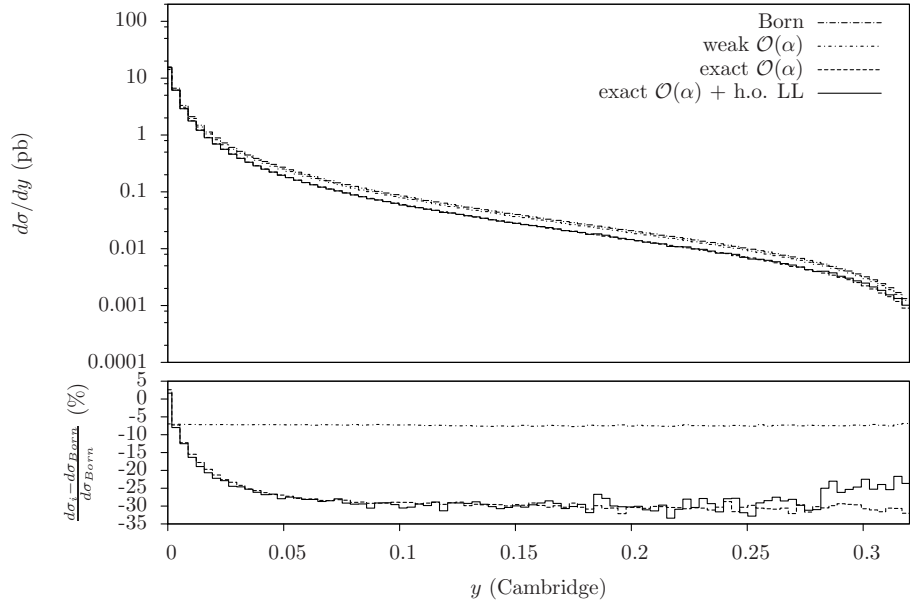}
\caption{Cambridge $y$ distribution at 350 GeV.}
\label{ycam-350}\end{center}\end{figure}

\begin{figure}\begin{center}
\includegraphics[width=12cm]{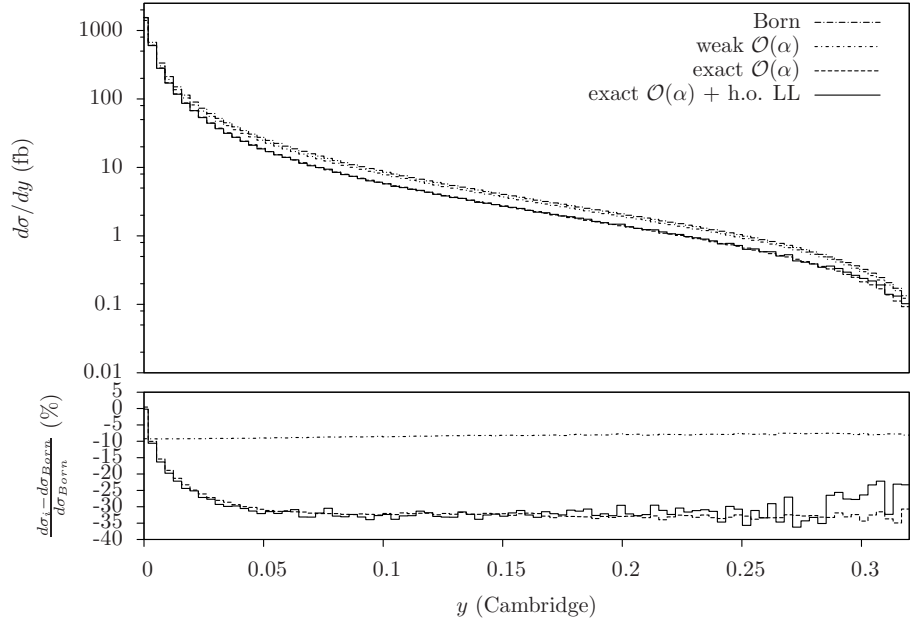}
\caption{Cambridge $y$ distribution at 1 TeV.}
\label{ycam-1000}\end{center}\end{figure}

In Figs.~\ref{ycam-peak}, \ref{ycam-350} and~\ref{ycam-1000}
the Cambridge $y$ distribution is shown. For each event, the observable 
$y$ is defined as the minimum (Cambridge) $y_{ij}$ such
as $y_{ij}> y_{cut} = 0.001$. Also on this distribution the weak
effects are quite constant and between --5\% and --10\% (increasing with
energy), while real radiation effects are significantly
larger (and negative) and distort the LO shape. 

\begin{figure}\begin{center}
\includegraphics[width=12cm]{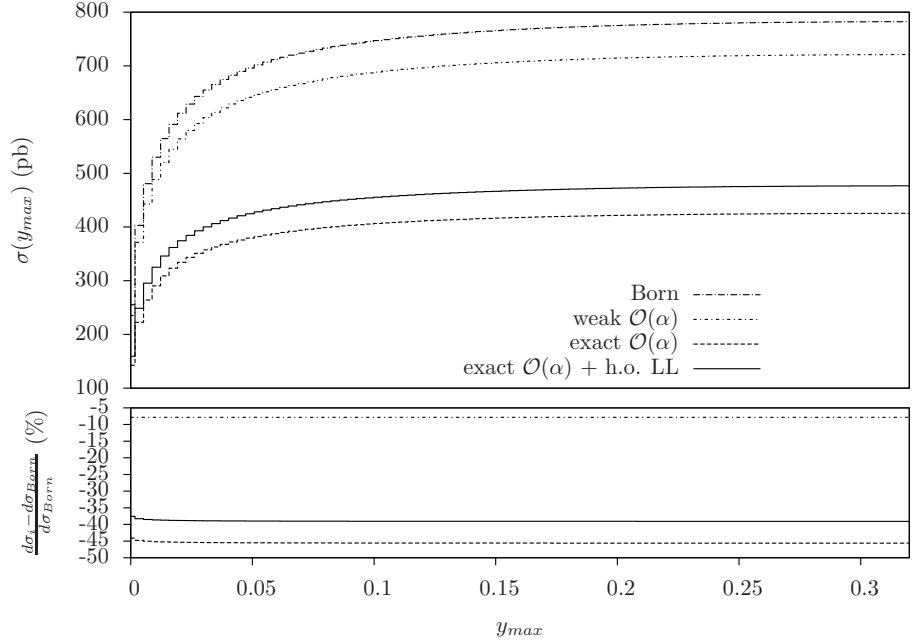}
\caption{Cross section as a function of the Cambridge maximum $y$ at the $Z$ peak.}
\label{ycamsum-peak}\end{center}\end{figure}

\begin{figure}\begin{center}
\includegraphics[width=12cm]{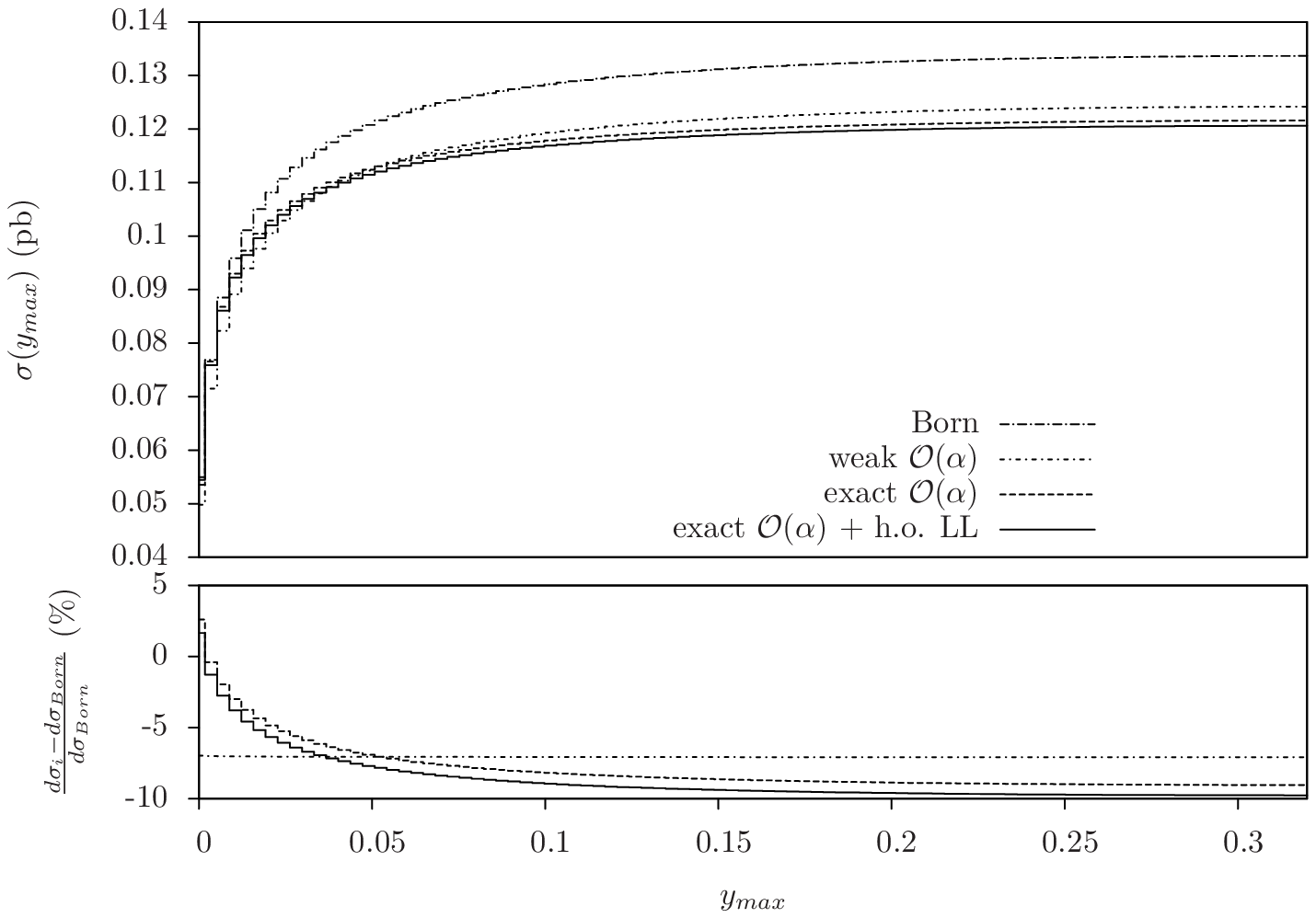}
\caption{Cross section as a function of the Cambridge maximum $y$ at 350 GeV.}
\label{ycamsum-350}\end{center}\end{figure}

\begin{figure}\begin{center}
\includegraphics[width=12cm]{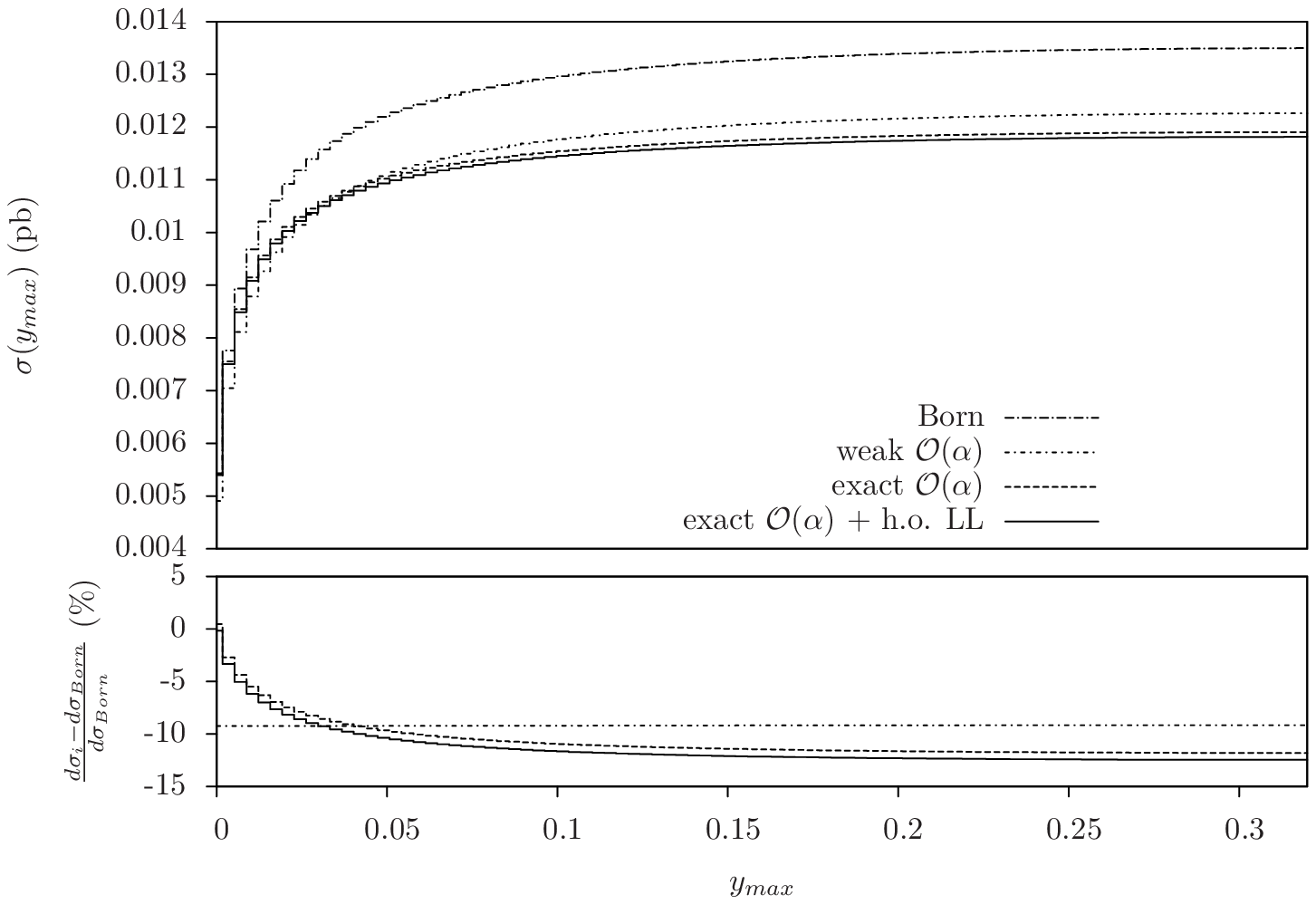}
\caption{Cross section as a function of the Cambridge maximum $y$ at 1 TeV.}
\label{ycamsum-1000}\end{center}\end{figure}

Figs.~\ref{ycamsum-peak}, \ref{ycamsum-350}
and~\ref{ycamsum-1000} present instead
the cross sections integrated over $y$ in the
range $y_{cut}< y < y_{max}$, as a function of $y_{max}$.
Corrections can be very large in such distributions, generally at any 
energy, reaching the minus several tens of percent 
(QED ones) or nearly the --10\% (weak ones) level.

If one combines $b$-(anti)quark flavour tagging with jet-charge measurements,
it is possible to define the forward-backward asymmetry, $A_{\rm FB}$, 
for the case of $b$-jet samples. As a clear anomaly in this observable 
(defined already at LO in the case of $b\bar b$ final states)
has
survived after the LEP and SLC era \cite{LEPEWWG,SLD}, it is worthwhile to investigate 
the impact that (hitherto neglected) EW contributions through 
${\cal O}(\alpha_{\rm S}\alpha_{\rm{EM}}^3)$ can potentially have 
in addressing the discrepancy between data and SM. Here, we use the
following definition for the asymmetry distribution as a function of
the $b\bar b$ invariant mass:
\begin{equation}
A_{\rm{FB}}(M_{b\bar
  b})=\frac{\int_0^{\pi/2}d\theta_b\frac{d\sigma}{d\theta_bdM_{b\bar
      b}} - \int_{\pi/2}^\pi
  d\theta_b\frac{d\sigma}{d\theta_bdM_{b\bar b}}}{\int_0^{\pi}
d\theta_b\frac{d\sigma}{d\theta_bdM_{b\bar b}}}
\label{afbM}
\end{equation}
with $\theta_b$
being the polar angle of the $b$-jet relative to the electron,
wherein both the numerator 
and denominator (normalisation) involve only terms of 
${\cal O}(\alpha_{\rm S}\alpha_{\rm{EM}}^2)$ plus
${\cal O}(\alpha_{\rm S}\alpha_{\rm{EM}}^3)$ and where
the final state comprises three jets. The integrals involved in
Eq.~(\ref{afbM}) are over the phase space allowed by the cuts defined in
the previous Section. Whilst the size of the
asymmetry is substantial through the computed orders, and the full
${\cal O}(\alpha_{\rm S}\alpha_{\rm{EM}}^3)$ rates are sizeably different
from the ${\cal O}(\alpha_{\rm S}\alpha_{\rm{EM}}^2)$ ones (at least
 at very small and very large energies), 
our results should clearly be folded with all known higher order corrections to the $e^+e^-\to
b\bar b$ process (work on this is in progress).

Finally, in Fig.~\ref{energyscan-afb}, the forward-backward asymmetry,
now integrated over $M_{b\bar b}$,
\begin{equation}
A_{\rm{FB}}=\frac{\int_0^{\pi/2}d\theta_b\frac{d\sigma}{d\theta_b} - \int_{\pi/2}^\pi
  d\theta_b\frac{d\sigma}{d\theta_b}}{\int_0^{\pi}
d\theta_b\frac{d\sigma}{d\theta_b}}
\label{afbsqrts}
\end{equation}
is plotted as a function of the CM energy and as obtained at lowest
order and by including only QED corrections or the full one-loop EW
corrections. As the CM energy raises, the increasing effect of the
purely weak corrections is more and more evident.


\section{Conclusions}
\label{Sec:Conclusions}

A careful analysis of actual $e^+e^-$ $\to$ $b\bar bg$ data 
is in order then, involving
 one-loop
EW effects. In this regard though, one {\it caveat} should be borne in mind: as emphasised in Refs.~\cite{Baur-rad,hadronic} (albeit in the hadronic
context), particular care should be devoted to the
treatment of real $W^\pm$ and $Z$ radiation  and decay in the
definition of the jet sample, as this will
determine whether tree-level $W^\pm$ and $Z$ bremsstrahlung
effects (neglected here) have to be included in the theoretical predictions
through ${\cal O}(\alpha_{\rm S}\alpha_{\rm{EM}}^3)$. However,
given the cleanliness of jet samples produced in electron-positron machines, as compared
to hadronic ones, we believe that the former contribution can
effectively be disentangled, thereby rendering our present predictions
of immediate experimental relevance. 

\section*{Acknowledgements} 
SM thanks FP for financial support during a visit to Pavia while
FP thanks SM and DAR for hospitality in
Southampton. CMCC acknowledges partial
 financial support from the British Council
in the form of a `Researcher Exchange Programme' award
and from the Royal Society via a `Short Visit to the UK' grant.




\begin{figure}\begin{center}
\includegraphics[width=12cm]{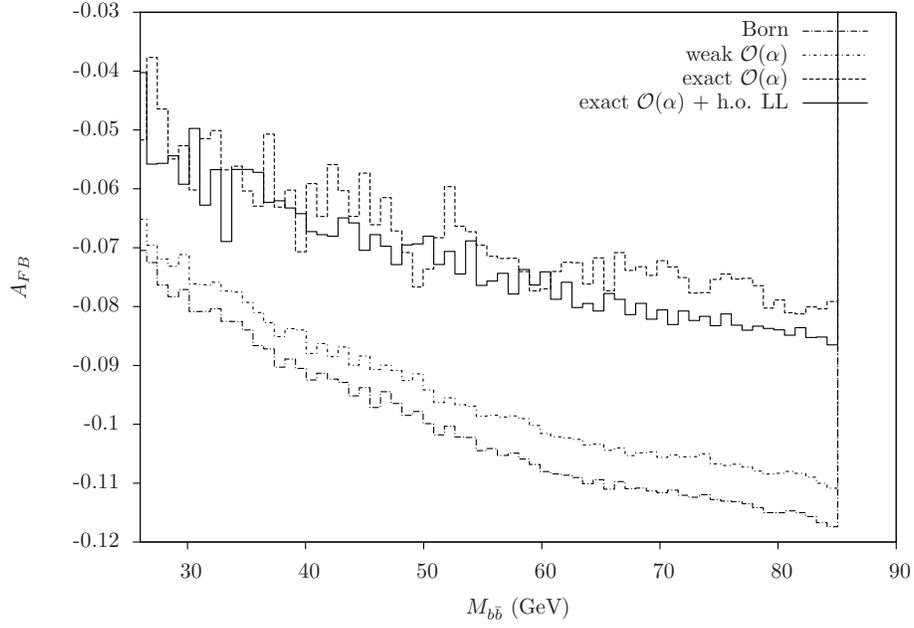}
\caption{$A_{\rm{FB}}$ distribution at the $Z$ peak.}
\label{afb-bbar-peak}\end{center}\end{figure}
\begin{figure}\begin{center}
\includegraphics[width=12cm]{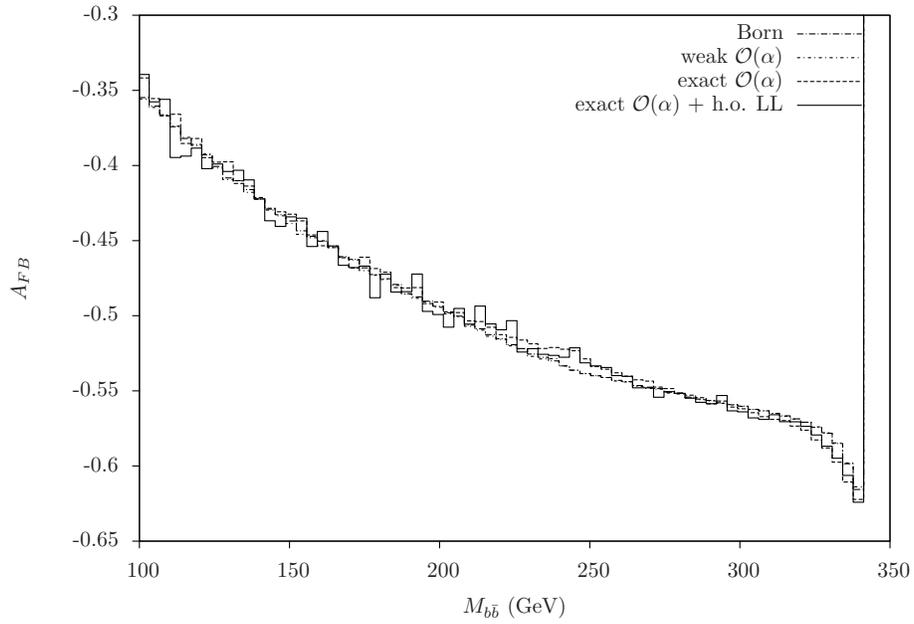}
\caption{$A_{\rm{FB}}$ distribution at 350 GeV.}
\label{afb-bbar-350}\end{center}\end{figure}
\begin{figure}\begin{center}
\includegraphics[width=12cm]{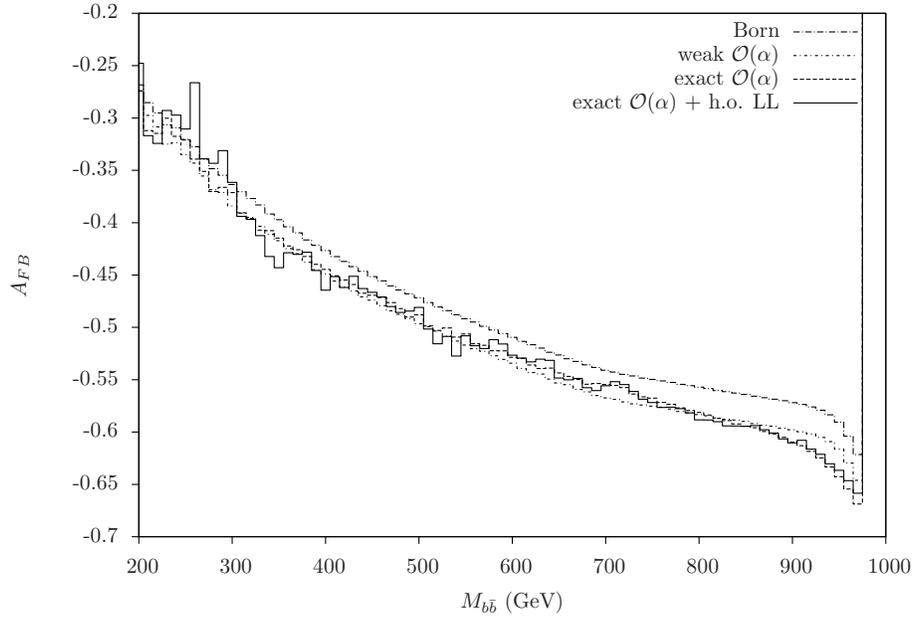}
\caption{$A_{\rm{FB}}$ distribution at 1 TeV.}
\label{afb-bbar-1000}\end{center}\end{figure}

\begin{figure}\begin{center}
\includegraphics[width=12cm]{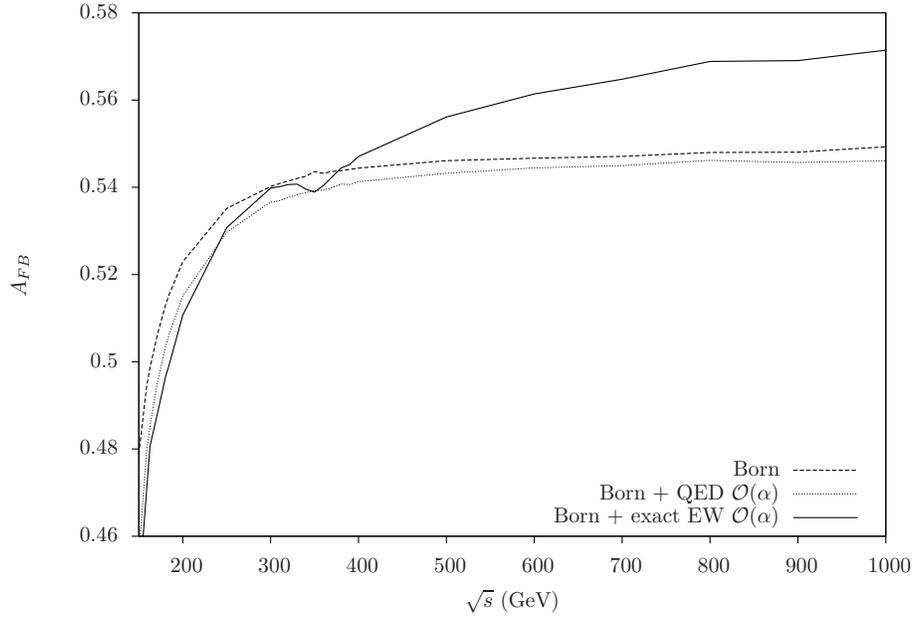}
\caption{$A_{\rm{FB}}$ as a function of the CM energy, obtained at
  tree level, adding only QED corrections and the full one-loop EW corrections.}
\label{energyscan-afb}\end{center}\end{figure}

\end{document}